# SINGLE-MODE PHOTONIC CRYSTAL FIBER WITH AN EFFECTIVE AREA OF 600μm² AND LOW BENDING LOSS

M.D. Nielsen[1,2], J.R. Folkenberg[1] and N.A. Mortensen[1]

[1] Crystal Fibre A/S
Blokken 84
DK-3460 Birkerød
Denmark

[2] COM
Technical University of Denmark
Building 345V
DK-2800 Kgs. Lyngby
Denmark

e-mail: mdn@crystal-fibre.com

*Abstract*: A single-mode all-silica photonic crystal fiber with an effective area of 600μm² and low bending loss is demonstrated. The fiber is characterized in terms of attenuation, chromatic dispersion and modal properties.

*Introduction*: The photonic crystal fiber (PCF) has since it was first proposed [1] attracted growing attention due to its many unique properties. One of the first special characteristics to be reported for the PCF was its potential to be endlessly single-mode (ESM) [2] referring to the absence of higher-order modes regardless of the optical wavelength. In the case of conventional fibers, the effective area is limited by the fact that an increasing core size requires a correspondingly decreasing index step between the core and the cladding in order to maintain single-mode operation. This imposes requirements on the control of the index profile which is difficult to realize with index-raising doping of the glass. A PCF which is ESM can in principle be scaled to an arbitrary dimension and remain single-mode. However, since the numerical aperture (NA) decreases with increasing mode size, the scaling of the PCF is in general limited by macro-bending loss and micro-deformation loss due to the decreasing mode spacing between the guided mode and leaky cladding-modes [3,4]. In contrast to conventional fibers, the bend-loss edge for the PCF is located at lower wavelengths compared to the transmission window [2] which is due to the fact that the mode spacing decreases with decreasing wavelength.

In solid core Bragg fibers, effective areas of more than 500 µm$^2$ at 1550 nm have been demonstrated but with attenuation levels in the order of dB/m [5]. PCFs with large mode-area have also been demonstrated [6,7] but only for structures which are intrinsically susceptible to loss caused by bending and other perturbations. In this work we present a single- mode PCF with an effective area of 600 µm$^2$ and low bending loss.

*Fiber design and properties*: The considered PCF is an all silica fiber with a number of air holes, of diameter d, running along the length of the fiber, placed on a triangular lattice arrangement with a pitch $\Lambda$. The central air hole in the structure is omitted creating a high-index defect allowing for guidance of light by total internal reflection. For the relative hole-size, d/$\Lambda$, below a value of 0.41 the propagation loss of the higher- order mode (HOM) increases drastically and it is not considered to be guided [8]. For d/$\Lambda$ larger than this value the propagation loss decreases and the structure now supports a HOM although the actual propagation loss might be significantly larger than that of the fundamental mode, causing the fiber in practice to act as a single-mode fiber. Pushing d/$\Lambda$ to larger values has the benefit of increasing the mode spacing and in order to investigate these properties, a number of fibers with a fixed cladding diameter of 280 µm and $\Lambda$=23 µm were fabricated, with d/Λ ranging from 0.45 to 0.53. All fibers were drawn from the same preform built using the stack and pull technique. The value of the cladding diameter was chosen to screen the waveguide from micro deformations [4]. The air-hole structure had 5 periods around the core and the fibers were coated with a single-layer actylate coating with a diameter of 426 μm.

*Characterization:* All fabricated fibers were characterized in terms of spectral attenuation from 1100 nm to 1700 nm using the cut-back technique. When characterized on a spool with a radius of 8 cm, all fibers showed varying degrees of bending induced attenuation at 1550 nm. However, on a spool of 16 cm radius the bend-loss edge of most fibers had moved to wavelengths lower than 1550 nm although fibers with d/Λ < 0.48 were still influenced by the tail of the bend loss edge. The fibers with the largest d/Λ value of 0.53 showed indication of supporting a high-order mode which was confirmed by inspecting the near field on a 50 cm sample. The fiber with d/Λ of 0.50 was single mode in the inspected wavelength region and not influenced by the bend-loss edge. In Fig. 1, the attenuation spectrum of this fiber, measured on a 150m sample, is shown. The steep bend loss edge is located at ~1200nm with a loss tail extending to the position of the OH absorption peak at 1.38 μm. The attenuation is 5 dB/km at 1550 nm at which is in good agreement with the recorded OTDR trace shown in the insert of Fig. 1. The low backscatter intensity observed in the OTDR measurement results from the low NA of the fiber which is estimated to be 0.04. The attenuation level at 1550 nm is mainly attributed to absorption caused by contamination introduced during the fabrication process such as OH and other impurities on the surfaces on the capillary tubes used to construct the preform. The attenuation level is similar to what is observed in fibers with significantly smaller $\Lambda$ when using the same fabrication process, and may be drastically improved by applying surface polish of the capillaries and dehydration techniques.

In Fig. 2, the measured chromatic dispersion from 1520 nm to 1640 nm (solid line) and the material dispersion of pure silica (dotted line) are shown along with the waveguide dispersion (dashed line) obtained from a numerical simulation using the plane-wave expansion method. The waveguide dispersion is almost constant with a value of 1 ps/(km·nm) and the chromatic dispersion is consequently dominated by the material dispersion with a value of 23.3 ps/(km·nm) at 1550 nm.

Based on the values of d/$\Lambda$ and $\Lambda$, the effective area, $A_{eff}$, can be simulated yielding a value of 591 µm² which corresponds to an equivalent Gaussian $1/e^2$ mode-field diameter (MFD) of 27.4 µm. In order to verify this experimentally, we study the near field at 1550 nm shown as a contour plot in the insert in Fig. 3 which also defines the two orthogonal directions x and y. In Fig. 3, the intensity profile of the near field along the x and y direction is shown (open circles and triangles, respectively). As seen from these profiles, the mode has a close to Gaussian shape which is indicated by the two fitted Gaussian functions (represented by the dashed and dotted line for the x and y direction, respectively). However, the near field is not rotational symmetric in the transverse plane but has a hexagonal shape leading to a difference in the $1/e^2$ MFD which can be extracted from the fits along the two directions. For the x direction, the MFD of the fitted Gaussian is 27.9 µm and for the y direction it is 29.8 µm. The $1/e^2$ width obtained directly from the raw data yields 25.5 µm and 29.2 µm for the x and y direction, respectively, corresponding to an average value of 27.2 µm. All measures of the MFD can thus be given as 27.7 +/- 8% which is in good agreement between the calculated value of $A_{eff}$. For $A_{eff}$ =600 µm$^2$, the nonlinear coefficient, $\gamma$=2$\pi$n$_2$/($\lambda A_{eff}$), yields a value of 0.16 (W·km)$^{-1}$, where $n_2$=2.4·10$^{-20}$ m$^2$/W is the nonlinear refractive index of pure silica.

*Conclusion:* We have demonstrated a PCF with an effective area of 600 µm$^2$ at 1550 nm by optimizing d/Λ to a value of 0.50 and thereby achieving both single-mode operation and low bending loss.

**Figure Captions:**

**Figure 1:**   Measured spectral attenuation from 1100 nm to 1700 nm yielding a value of 5 dB/km at 1550 nm. The insert shows an OTDR trace at 1550 nm.

**Figure 2:**   Measured chromatic dispersion (solid line), material dispersion of pure silica (dotted line) and calculated waveguide dispersion (dashed line) from 1520 nm to 1640 nm.

**Figure 3:**   Insert shows contour plot of measured near field profile and definition of the directions x and y. The curves show intensity profiles along the directions x and y (circles and squares, respectively) along with fitted Gaussian profiles (solid and dashed lines respectively).

**Figure1**

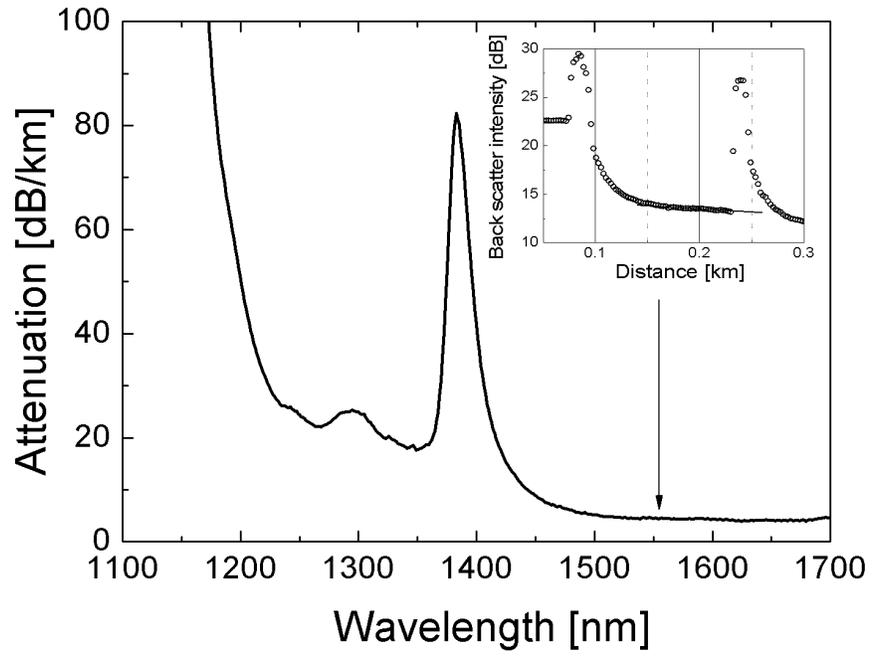

**Figure 2**

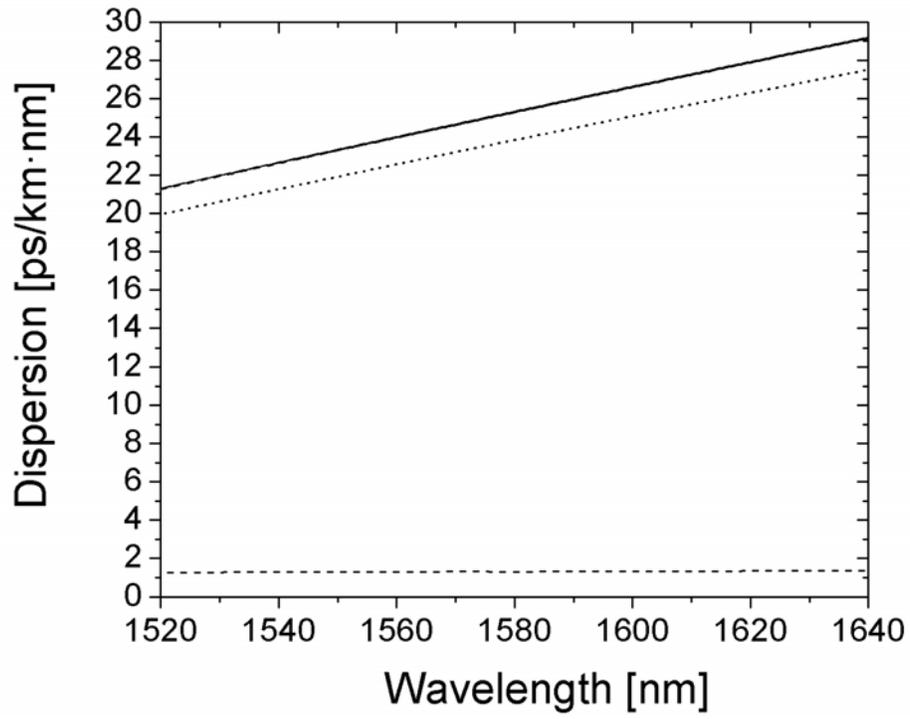

**Figure 3**

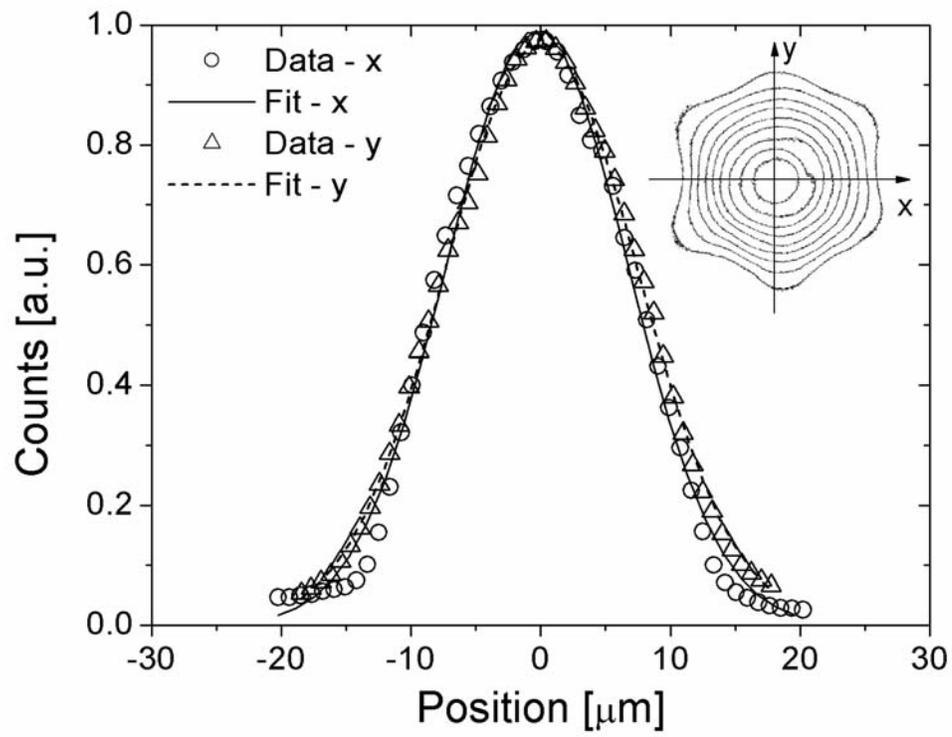